\documentclass[aps,prb,twocolumn,showpacs,groupedaddress]{revtex4}

\usepackage{graphicx,bm}
\usepackage{amsmath,amssymb}
\usepackage{hyperref}
\usepackage{graphicx,amsfonts,amsbsy}

\newcommand{\D}{\Delta}

\newcommand{\G}{\Gamma}
\renewcommand{\L}{\Lambda}

\newcommand{\W}{\Omega}

\renewcommand{\c}{\chi}
\renewcommand{\d}{\delta}
\newcommand{\e}{\epsilon}
\newcommand{\f}{\phi}

\renewcommand{\l}{\lambda}
\newcommand{\m}{\mu}
\newcommand{\n}{\nu}
\newcommand{\s}{\sigma}
\renewcommand{\r}{\rho}
\newcommand{\p}{\pi}
\newcommand{\q}{\theta}

\newcommand{\vc}[1]{\mathbf{#1}}
\newcommand{\ua}{\uparrow}
\newcommand{\da}{\downarrow}
\renewcommand{\pl}{\parallel}
\newcommand{\tl}[1]{\tilde{#1}}

\newcommand{\ord}{\mathcal{O}}
\newcommand{\bra}[1]{\langle #1 |}
\newcommand{\ket}[1]{| #1 \rangle}

\newcommand{\comments}[1]{}

\begin{document}

\title{Bilayer quantum Hall system at $\nu_t = 1$: pseudospin models and in-plane magnetic field}

\author{O. Tieleman}
\email{o.tieleman@uu.nl}

\affiliation{Institute for Theoretical Physics, Utrecht
University, Leuvenlaan 4, 3584 CE Utrecht, The Netherlands}

\author{A. Lazarides}

\affiliation{Institute for Theoretical Physics, Utrecht
University, Leuvenlaan 4, 3584 CE Utrecht, The Netherlands}

\author{D. Makogon}

\affiliation{Institute for Theoretical Physics, Utrecht
University, Leuvenlaan 4, 3584 CE Utrecht, The Netherlands}

\author{C. Morais Smith}

\affiliation{Institute for Theoretical Physics, Utrecht
University, Leuvenlaan 4, 3584 CE Utrecht, The Netherlands}

\date{\today}


\begin{abstract}
We investigate two theoretical pseudomagnon-based models for a bilayer quantum Hall system (BQHS) at total filling factor $\nu_t = 1$. We find a unifying framework which elucidates the different approximations that are made. We also consider the effect of an in-plane magnetic field in BQHSs at $\nu_t = 1$, by deriving an equation for the ground state energy from the underlying microscopic physics. Although this equation is derived for small in-plane fields, its predictions agree with recent experimental findings at stronger in-plane fields, for low electron densities. We also take into account finite-temperature effects by means of a renormalisation group analysis, and find that they are small at the temperatures that were investigated experimentally.
\end{abstract}
\vskip2pc

\maketitle

\section{Introduction}

Over the last two decades, numerous experiments have been performed on bilayer quantum Hall systems (BQHSs) at total filling factor $\n_t =1$, revealing a very rich physical system. In a series of groundbreaking experiments in the early years of the current decade, evidence was found for the existence of an excitonic superfluid,\cite{ke04, ke02, tu03} as well as of the associated Goldstone mode.\cite{sp01} This phase is destroyed by increasing the effective interlayer separation $d / l$, where $d$ is the distance between the layers and $l$ the magnetic length. For small effective separations, the system is in an incompressible phase and the Hall conductivity exhibits a plateau at $\n = 1$, whereas at large effective separations, the layers decouple and the Fermi-liquid behaviour of two independent layers is recovered; here, the system is in a compressible phase. The critical effective interlayer separation $(d / l)_c$ has been shown to be sensitive to charge imbalance, tunnelling amplitudes and in-plane magnetic fields.\cite{ch08} In addition to the compressible-incompressible  transition discussed above, a commensurate-incommensurate transition has been identified, driven by an in-plane magnetic field.\cite{mu94,ya94} Recently, detailed measurements on this aspect of the system~\cite{fu08} have become available.

This wealth of experimental findings has naturally renewed the theoretical interest in the system. At present, three independent models of the neutral (spin-flip) excitations of a bilayer system at $\n_t = 1$ and the case of equal electron populations in each layer exist in the literature,\cite{fe89, ma90, do06} with three different derivations and two different predicted spectra. This state of affairs merits an investigation. That is the first aim of this paper: to compare and contrast, and where possible to link, the existing models. Random phase approximation (RPA) calculations have revealed a linear Goldstone mode,\cite{fe89} in qualitative and even rough quantitative agreement with experiment.\cite{sp01} Subsequently, an approximation based on pseudospin waves has been proposed for the system in the presence of strong tunnelling,\cite{ma90} which reproduces the RPA result from Ref.~\onlinecite{fe89} in the non-tunnelling limit. Lastly, in recent years, a bosonisation method has been proposed\cite{do06} to directly study the Bose-Einstein condensate (BEC) of excitons detected experimentally.~\cite{ke04, ke02, tu03, sp01} The same $\nu_T=1$ system with \emph{unequal} layer electron populations has also been studied using similar techniques in Ref.~\onlinecite{ya01}. We will develop a unifying framework that links all three models for the balanced case and allows for further approximations.

This first aim is subsidiary to a second purpose: to create a model for a bilayer system at $\n_t = 1$ with an in-plane magnetic field $\vc{B}_\pl$. Apart from the above-mentioned experimental findings, in-plane magnetic fields are often used to suppress tunnelling between the layers, in order to study the quantum Hall effect (QHE) in bilayer systems without interlayer tunnelling.\cite{mu94, ka09}

Early theoretical work in this area has led to the prediction of a commensurate-incommensurate phase transition.\cite{ya94,ya96} We will revisit the problem, and derive an equation for the ground state energy from the underlying microscopic physics. This equation was proposed in Ref.~\onlinecite{ya94}, but is explicitly derived here. Following the work by Hanna \emph{et al.}\cite{ha01} we calculate at which in-plane magnetic field strength the commensurate-incommensurate transition should occur, with the aim of comparing the theoretical prediction to recent experimental observations.\cite{fu08} Although this model is based on small in-plane fields, it turns out to agree with experiments in the regime of low electron densities. Including finite-temperature effects by means of a renormalisation group analysis (details of which are presented elsewhere\cite{alip}) yields a small change in critical in-plane field, which is not enough to explain the difference between the theoretical predictions and the experimental observations at larger electron densities.

This paper is structured as follows. In section II, we introduce the BQHS and derive its microscopic Hamiltonian. In section III, we discuss the various models that exist in the literature, introduce one new bosonisation approach, and analyse the differences between the models. This analysis results in a unifying framework where all the models that are investigated here can be seen to be variations of each other. In section IV, we introduce an in-plane field into the bilayer system, and microscopically derive an equation for the ground state energy of the bilayer system with in-plane field. We then present some results coming out of that model, including a finite-temperature result obtained by means of a renormalisation group analysis. Section V contains conclusions and a discussion.

\section{The system}\label{secsystem}

A BQHS consists of two individual two-dimensional electron gas layers, parallel and at a distance $d$ to each other. The QHE occurs in such systems at low temperatures and under strong perpendicular magnetic fields. In these circumstances, the system is characterised by four parameters: the total \emph{filling factor} $\n_t$, the interlayer separation $d$, the tunnelling amplitude $t$, and the charge imbalance $\D \n$. The total filling factor is given by the sum of the filling factors of the upper ($\n_u$) and lower ($\n_l$) layers, $\n_t = \n_u + \n_l$\comments{  $ = n_e / n_\f$ counts the number of completely filled Landau levels}, whereas the charge imbalance $\D \n = \n_u - \n_l$ is the difference between the individual layer filling factors. The filling factor of an individual layer $\n = n_e / n_\f$ counts the number of filled Landau levels within that layer. We will concentrate on a system with $\n_t = 1$ and zero charge imbalance, such that each layer has filling factor 1/2. This restricts the dynamics to the lowest Landau level (LLL), which means that we must project all operators into the LLL. The other two important parameters are the effective interlayer separation $d / l$, where $l = \sqrt{\hbar / e B}$ is the magnetic length, and the ratio between the tunnelling amplitude $t$ and the characteristic Coulomb interaction energy $E_{\rm C} = e^2 / \e l$. These two dimensionless parameters can be tuned experimentally through the magnetic field strength $B$. By varying $d / l$, the relative importance of the interlayer Coulomb interaction is changed, since $l$ is proportional to the average distance between two neighbouring electrons within one layer. Increasing $t / E_{\rm C}$ enhances the interlayer coherence, since the tunnelling energy favours a coherent state.

We will take the temperature and Zeeman splitting to be such that the electron spins are completely frozen. From a single-particle viewpoint, the layer degree of freedom gives rise to a two-state system, so the relevant Hilbert space is the same as that of a spin-1/2 system. Using this similarity, we map the system to a single-layer system where the layer degree of freedom is described by a pseudospin variable. In the absence of tunnelling or a bias voltage, there is no energy associated with the pseudospin of an electron. The Coulomb interaction, however, \emph{is} pseudospin-dependent, since the interlayer repulsion is weaker than the intralayer repulsion.

Orienting the axes of the spin\footnote{Since the real electron spins do not play a role in the subsequent discussion, we will drop the prefix `pseudo' and refer to pseudospin as `spin'.} system such that $\ua$ ($\da$) indicates an electron in a(n) (anti)symmetric superposition of being in both layers, the Coulomb part of the Hamiltonian takes the form
\begin{align}
H_{\rm C} = \frac{1}{2} \sum_{\s, \s', \vc{k}} v_{\s, \s'}(\vc{k}) \r_{\s, -\vc{k}} \r_{\s', \vc{k}}.
\end{align}
where $v_{\s, \s'}(\vc{k})$ is the spin-dependent Coulomb interaction. The density operator $\r_{\s, \vc{k}}$ is given by
\begin{align}
\label{eldensityop}
\r_{\s, \vc{k}} = e^{-|l\vc{k}|^2 / 2} \sum_{m, n} G_{m, n}(\vc{k}) c^\dag_{\s, m} c_{\s, n},
\end{align}
where the operator $c^{(\dag)}_{\s, m}$ destroys (creates) an electron with spin $\s$ in the guiding centre $m$. The function $G_{m, n}$ is defined in Appendix \ref{lllprojectionalg}. Similarly, we can define spin density operators for later use:
\begin{subequations}
\label{spindensityops}
\begin{align}
S^z_\vc{k} = & \, \frac{e^{-|l\vc{k}|^2 / 2}}{2} \sum_{m, n} G_{m, n}(\vc{k}) \bigl( c^\dag_{\ua, m} c_{\ua, n} - c^\dag_{\da, m} c_{\da, n} \bigr), \\
S^x_\vc{k} = & \, \frac{e^{-|l\vc{k}|^2 / 2}}{2} \sum_{m, n} G_{m, n}(\vc{k}) \bigl( c^\dag_{\ua, m} c_{\da, n} + c^\dag_{\da, m} c_{\ua, n} \bigr).
\end{align}
\end{subequations}
The density operators defined in Eqs.~\eqref{eldensityop} and \eqref{spindensityops} obey the LLL projection algebra, which is discussed in Appendix \ref{lllprojectionalg}. The Coulomb Hamiltonian can be split up into total density and spin parts by writing
\begin{align}
H_C = & \, \frac{1}{2} \sum_\vc{k} v_0(\vc{k}) \r_{-\vc{k}} \r_\vc{k} + 2 \sum_\vc{k} v_c(\vc{k}) S^x_{-\vc{k}} S^x_\vc{k},
\end{align}
where
\begin{align}
v_{0/c}(\vc{k}) = \frac{\p e^2}{\e |\vc{k}|} \bigl(1 \pm e^{- |\vc{k}| d} \bigr)
\end{align}
(note that we are working in a unit area system: $A = 1$). The $v_c$-term measures the capacitive energy due to charge imbalance between the layers, since $S^x$ is proportional to the charge imbalance. Interlayer tunnelling adds a term $H_T$:
\begin{align}
H_T = & \, - t \int d^2 r \sum_{m, n} \frac{e^{-|\vc{r}|^2 / 2 l^2}}{2 \p l^2} G_{m, n}(\vc{r}) \Bigl( c^\dag_{u, m} c_{l, n} + c^\dag_{l, m} c_{u, n} \Bigr) \notag \\
= & \, - t S^z_{\vc{k} = 0},
\end{align}
where the relation between $c_{u/l}$ and $c_{\ua, \da}$ is given by $c_{\ua/\da} = 2^{-1/2} (c_u \pm c_l)$. Since the kinetic term is constant and therefore irrelevant in the $\n_t = 1$ case, we now have the total Hamiltonian
\begin{align}\label{spinham}
H = - t S^z_\vc{0} + \frac{1}{2} \sum_\vc{k} v_0(\vc{k}) \r_{-\vc{k}} \r_\vc{k} + 2 \sum_\vc{k} v_c(\vc{k}) S^x_{-\vc{k}} S^x_\vc{k}.
\end{align}
The $v_0$-term is invariant under rotations of the spin, whereas the $v_c$-term favours $S^x = 0$ due to the positivity of $v_c$. In other words, it disfavours spin orientations that do not lie in the plane parallel to the layers. In the absence of the tunnelling term, the spin orientation is constrained to lie in the plane but is otherwise free.

Note that the orientation of the spin axes can be chosen freely: for example, one can choose to use spin up (down) to represent an electron in the upper (lower) layer. In that case, the roles of $S^x$ and $S^z$ in Eq.~\eqref{spinham} are inverted and it is the spin $z$-axis that is perpendicular to the layers.

\section{Pseudospin models}

To accomodate the presence of a strong tunnelling term, we will work in the symmetric/antisymmetric (S/AS) basis. Starting from a symmetric ground state, we study the antisymmetric excitations, which are approximately bosonic in nature. Defining the bosonic vacuum to be the ferromagnet, we describe the excitations above the ferromagnet as a system of noninteracting bosons.

\subsection{Magnons}
In the presence of tunnelling, the symmetric state has the lowest energy, since the single-electron wavefunction for that state has no nodes in the direction perpendicular to the layers, whereas the one for the antisymmetric state has a single node. The level splitting between the symmetric and antisymmetric states is proportional to the tunnelling amplitude and allows us to represent the ground state as a ferromagnet, in which all spins point in the same direction. This implies that a particular direction in spin space is selected. It is the tunnelling term that determines the preferred direction and thus breaks the in-plane symmetry. In our chosen spin orientation, its momentum space form is $H_T = - t S^z_\vc{0}$, which favours a uniform state in which the spin is oriented in the spin $z$ direction everywhere, parallel to the plane of the layers.

We have chosen the spin orientation such that the resulting Hamiltonian is diagonal when $v_c=0$, which occurs when the interlayer distance vanishes. In that case, the ground state, which we denote by $\ket{\chi}$, is the state in which all spins are oriented along the spin $z$-axis: $\ket{\chi} = \ket{ \hspace{-3pt} \ua\ua\ldots\ua}$\comments{; here, $\ket{\hspace{-2pt} \ua}$ is a single-particle state which satisfies $S^z \ket{\hspace{-2pt}\ua} = \ket{\ua}$}. If, on the other hand, $v_c \neq 0$, the Hamiltonian is no longer diagonal, because $\ket{\chi}$ is not an eigenstate of $S^x$.

The excitations above this ferromagnet are magnons, created by the operator $S^-_\vc{k} = S^x_\vc{k} - i S^y_\vc{k}$. With the help of the LLL projection algebra (Appendix \ref{lllprojectionalg}), it is easy to check that $\bra{\c}[S^-_\vc{q}, S^+_{-\vc{p}}] \ket{\c} \propto \d_{\vc{p}, \vc{q}}$. This implies that near the ground state, modes created by $S^-$ are approximately bosonic. Following earlier work,\cite{ma90, gi86} we can define a magnon operator $m^\dag$ by normalising $S^-$,
\begin{align}
m^\dag_\vc{k} = \frac{e^{|l\vc{k}|^2 / 4}}{\sqrt{N}} S^-_\vc{k},
\end{align}
where $N$ is the total number of electrons in the system. By explicit calculation, the single-magnon modes $\ket{\vc{k}} = m^\dag_\vc{k} \ket{\c}$ can be seen to be exact orthonormal eigenstates of the $v_c=0$ Hamiltonian. Hence, our choice of spin orientation is appropriate to study excitations above the small-$v_c$ limit of the system.

\subsection{Models}\label{secmodels}

\subsubsection{Single-mode approximation}\label{secsma}
MacDonald \emph{et al.} have proposed a model based on the magnon-like excitations mentioned above.\cite{ma90} Here, we briefly repeat the main results of that study with the aim of comparing them to the outcomes of other studies, as well as our own findings. The calculation of the excitation spectrum is based on a tried and tested method:\cite{la80} assuming the magnon density is low enough for the interaction to be negligible, one takes
\begin{align}
\e_\vc{k} = \bra{\vc{k}} H \ket{\vc{k}} - \bra{\c} H \ket{\c}.
\end{align}
Ignoring the magnon-magnon interaction comes down to computing the excitation spectrum as if there is only one mode; hence, it is called the \emph{single-mode approximation}. Taking into account the off-diagonal elements of the Hamiltonian in the magnon basis, one obtains
\begin{align}\label{smaham}
H = E_0 + \frac{1}{2} \sum_\vc{k} \Bigl[ & \, \e^{\rm sma}_\vc{k} m^\dag_\vc{k} m_\vc{k} \\
& \, + \frac{\l^{\rm sma}_\vc{k}}{2} \bigl( m^\dag_\vc{k} m^\dag_{-\vc{k}} + m_\vc{k} m_{-\vc{k}} \bigr) \Bigr], \notag
\end{align}
where
\begin{align}
\e^{\rm{sma}}_\vc{k} = & \, t + N e^{-|l\vc{k}|^2 / 2} v_c(\vc{k}) \\
& \, + \sum_\vc{q} e^{-|l\vc{q}|^2 / 2} \Bigl[ v_c(\vc{q}) + 2 v_0(\vc{q}) \sin^2(\vc{k} \wedge \vc{q} / 2) \Bigr], \\
\l^{\rm{sma}}_\vc{k} = & \, N e^{-|l\vc{k}|^2 / 2} v_c(\vc{k}) + \sum_\vc{q} v_c(\vc{q}) e^{-|l\vc{q}|^2 / 2} \cos(\vc{k} \wedge \vc{q}), \notag
\end{align}
and we have used the shorthand notation $\vc{k} \wedge \vc{q} = l^2 \hat{z} \cdot (\vc{k} \times \vc{q})$. Diagonalising Eq.~\eqref{smaham} by means of a Bogolyubov transformation one obtains
\begin{align}
H = \frac{1}{2} \sum_\vc{k} \W^{\rm sma}_\vc{k} a^\dag_\vc{k} a_\vc{k},
\end{align}
where $a_\vc{k}$ are the quasiparticle excitations. The Bogolyubov spectrum has the familiar form
\begin{align}
\W^{\rm sma}_\vc{k} = \sqrt{(\e^{\rm sma}_\vc{k})^2 - (\l^{\rm sma}_\vc{k})^2}.
\end{align}
Taking the limit $t \to 0$, $\W^{\rm sma}_\vc{k}$ reduces to the spectrum found within the RPA formalism.~\cite{fe89} Comparing the above calculation of the spectrum to the one performed in Ref.~\onlinecite{fe89}, one sees that up to a minus sign, the same pseudospin mapping is used. The only difference is the manner of calculation: the former uses Feynman diagrams, whereas the latter uses the approximately bosonic nature of the pseudospin waves. The predicted spectrum features a Goldstone mode at $t = 0$, as found experimentally,\cite{sp01} but also a phase transition at $(d / l)_c \approx 1.2$, whereas experiments reveal the critical separation to lie at $(d / l)_c \approx 1.8$. Furthermore, the theoretically predicted phase transition is induced by a roton minimum touching the axis, whereas no roton has ever been observed, in spite of an extensive search.\cite{eispriv}

\subsubsection{S/AS bosonisation}\label{secsasbos}
Recently, another approach has been presented, called \emph{bosonisation}, which is also based on a pseudospin mapping and bosonic excitations.\cite{do06} However, a different spin orientation is used: spin up (down) refers to an electron being in the upper (lower) layer instead of an (anti)symmetric state. This spin mapping is useful, since it allows one to describe very directly the BEC of excitons that has been observed experimentally.~\cite{ke04, ke02, tu03, sp01} However, after the bosonisation procedure, the Hamiltonian describes a system in which every second electron is part of a boson. This is problematic, because the bosonic operators were defined under the assumption that there would be few bosons: so few that they would locally only see the (ferromagnetic) ground state.

Here, we will present a bosonisation scheme based on the S/AS splitting discussed above, but still within the spin mapping used in Ref.~\onlinecite{do06}. In this way, we end up with a system containing few bosons, maintaining the validity of the approximation, but the bosons being described are no longer the excitons observed in Refs.~\onlinecite{sp01, ke02, ke04, tu03}.

The ground state $\ket{\c}$ is given by $\prod_m c^\dag_{\ua,m} \ket{0}$, and antisymmetric excitations are created by the operator
\begin{align}
R^-_\vc{k} = e^{-|l\vc{k}|^2 / 2} \sum_{m, n} G_{m, n}(\vc{k}) c^\dag_{\da, m} c_{\ua, n}.
\end{align}
Normalising $R^-$ in order to define a proper bosonic operator, we find
\begin{align}
b^\dag_\vc{k} = \frac{e^{|l\vc{k}|^2 / 4}}{\sqrt{N}} R^-_\vc{k}.
\end{align}
Note that $R^-_\vc{k}$ is identical to $S^-_\vc{k}$. Now, following Ref.~\onlinecite{do04}, we find bosonic expressions for the operators $\r$ and $S^{x/z}$, which appear in the fermionic Hamiltonian Eq.~\eqref{spinham}. To obtain these expressions, we first calculate the commutators of the operators $\r$ and $S^z$ with $b^\dag$. We find
\begin{align} \label{eqcommbdag}
\begin{split}
[\r_\vc{k}, b^\dag_\vc{q}] = & \, 2 \, i \, e^{-|l\vc{k}|^2 / 4} \sin(\vc{k} \wedge \vc{q} / 2) b^\dag_{\vc{q} + \vc{k}}, \\
[S^z_\vc{k}, b^\dag_\vc{q}] = & \, e^{-|l\vc{k}|^2 / 4} \cos(\vc{k} \wedge \vc{q} / 2) b^\dag_{\vc{q} + \vc{k}}.
\end{split}
\end{align}
With the commutators from Eq.~\eqref{eqcommbdag} and the actions of $\r$ and $S^z$ on the ground state, we find the following bosonic expressions for $\r$ and $S^z$ by means of the method outlined in Ref.~\onlinecite{do04}:
\begin{align}
\r_\vc{k} = & \, N \d_{\vc{k}, 0} + 2 \, i \, e^{-|l\vc{k}|^2 / 4} \sum_\vc{q} \sin(\vc{k} \wedge \vc{q} / 2) b^\dag_{\vc{k} + \vc{q}} b_\vc{q}, \notag \\
S^z_\vc{k} = & \, \frac{N \d_{\vc{k}, 0}}{2} - e^{-|l\vc{k}|^2 / 4} \sum_\vc{q} \cos(\vc{k} \wedge \vc{q} / 2) b^\dag_{\vc{k} + \vc{q}} b_\vc{q}.
\end{align}
To obtain a bosonic expression for $S^x$, we note that $S^x = (R^+ + R^-) / 2$, and simply invert the definition of $b^\dag$. Inserting the bosonic expressions for $\r$ and $S^{x/z}$ into the fermionic Hamiltonian yields a quadratic, but non-diagonal bosonic Hamiltonian
\begin{align}
H = \frac{1}{2} \sum_\vc{k} \Bigl[ \e_\vc{k}^{\rm bos} b^\dag_\vc{k} b_\vc{k} + \frac{\l_\vc{k}^{\rm bos}}{2} \bigl( b^\dag_\vc{k} b^\dag_\vc{-k} + b_\vc{k} b_\vc{-k} \bigr) \Bigr].
\end{align}
Again performing a Bogolyubov transformation to diagonalise the Hamiltonian, we find a quasiparticle spectrum
\begin{align}
\W^{\rm bos}_\vc{k} = & \, \sqrt{(\e_\vc{k}^{\rm{bos}})^2 - (\l_\vc{k}^{\rm{bos}})^2},
\end{align}
where
\begin{subequations} \label{eqbosoresult}
\begin{align}
\l_\vc{k}^{\rm{bos}} = & \, N e^{-|l\vc{k}|^2 / 2} v_c(\vc{k}) \quad \mbox{and} \\
\begin{split}
\e_\vc{k}^{\rm{bos}} = & \, t + N e^{-|l\vc{k}|^2 / 2} v_c(\vc{k}) + \\
& \, 2 \sum_\vc{q} e^{-|l\vc{q}|^2 / 2} v_0(\vc{q}) \sin^2(\vc{k} \wedge \vc{q} / 2).
\end{split}
\end{align}
\end{subequations}
This result is similar to that of Sec.~\ref{secsma}, but not identical, even though the same excitations were studied, and the same physical property (the approximately bosonic nature of the excitations) was the starting point for the approximation. In the following section, we explore the origin of this unexpected difference.

\subsection{Unification}

In the two approaches outlined above, the bosonic excitations being studied are the same (both are antisymmetric excitations above a symmetric ground state), and the method is the same (both are based on a pseudospin mapping), yet they yield different results. The origin of this apparent inconsistency can be found by expanding the Hamiltonian in $n$-magnon states:
\begin{align} \label{magnonexpansion}
H = & \, \sum_{n = 0}^\infty \sum_{\vc{q}_1, \dots, \vc{q}_n} \sum_{m = 0}^n \ket{\vc{q}_1, \dots, \vc{q}_m} \times \\
& \, \bra{\vc{q}_1, \dots, \vc{q}_m} H \ket{\vc{q}_{m+1}, \dots, \vc{q}_n} \bra{\vc{q}_{m+1}, \dots, \vc{q}_n}. \notag
\end{align}
Having assumed a low enough magnon density for the magnon-magnon interaction to be negligible, we need only take this series up to $n = 2$. From the truncated series, let us investigate the term $\sum_{\vc{p}, \vc{q}} \ket{\vc{p}} \bra{\vc{p}} H \ket{\vc{q}} \bra{\vc{q}}$, or more specifically, the $v_c$-term
\begin{align*}
2 & \, \sum_{\vc{k}, \vc{p}, \vc{q}} v_c(\vc{k}) \bra{\vc{p}} S^x_{-\vc{k}} S^x_\vc{k} \ket{\vc{q}} \\
= \frac{1}{2} & \, \sum_{\vc{k}, \vc{p}, \vc{q}} v_c(\vc{k}) \bra{\c} b_\vc{p} (S^+_{-\vc{k}} + S^-_{-\vc{k}})(S^+_\vc{k} + S^-_\vc{k}) b^\dag_\vc{q} \ket{\c}.
\end{align*}
Expanding the brackets yields four terms, of which the one of interest is
\begin{align*}
\frac{1}{2} \sum_{\vc{k}, \vc{p}, \vc{q}} \frac{e^{-|l\vc{k}|^2 / 2}}{N} v_c(\vc{k}) \bra{\c} b_\vc{p} b_\vc{k} b^\dag_\vc{k} b^\dag_\vc{q} \ket{\c}.
\end{align*}
Calculating the ground state expectation value $\bra{\c} b_\vc{p} b_\vc{k} b^\dag_\vc{k} b^\dag_\vc{q} \ket{\c}$ by means of the LLL projection algebra (i.e. commuting $S^\pm$) yields
\begin{align*}
\frac{1}{2} & \, \sum_{\vc{k}, \vc{p}, \vc{q}} \frac{e^{-|l\vc{k}|^2 / 2}}{N} v_c(\vc{k}) \bra{\c} [[b_\vc{p}, [b_\vc{k}, b^\dag_\vc{k}]], b^\dag_\vc{q}] \ket{\c} \\
= \frac{1}{2} & \, \sum_{\vc{k}, \vc{p}, \vc{q}} \frac{e^{-|l\vc{k}|^2 / 2}}{N} v_c(\vc{k}) \bra{\c} [b_\vc{p}, b^\dag_\vc{q}] \ket{\c} \\
= \frac{1}{2} & \, \sum_{\vc{k}, \vc{p}, \vc{q}} e^{-|l\vc{k}|^2 / 2} \d_{\vc{p}, \vc{q}} v_c(\vc{k}),
\end{align*}
which is precisely the difference between the two approaches. In the bosonisation scheme, the commutator $[b_\vc{k}, b^\dag_\vc{k}]$ is replaced with its ground state expectation value, which is a number. Hence, the outer commutators vanish, and the term does not enter the spectrum. A similar explanation holds for the difference between $\l^{\rm bos}_\vc{k}$ and $\l^{\rm sma}_\vc{k}$.

In short: taking the series in Eq.~\eqref{magnonexpansion} up to $n = 2$, we recover the results found by MacDonald \emph{et al.}\cite{ma90} The difference between the bosonisation scheme presented above and the one employed in Ref.~\onlinecite{ma90} arises from the fact that the former approximates the state of the system by the ground state already when evaluating the series in Eq.~\eqref{magnonexpansion}, whereas in the latter, this assumption is only made at the stage of the  Bogolyubov transformation.

It should be noted that the difference between the two approaches has nothing to do with the different spin representations used. They can be transformed into each other by rotating the spins, but such a rotation does not have any effect on the physical quantities that can be calculated from the models. The representations are only different in the way in which they describe the physics, but not in the physical approximation that is made. The difference in results between sections \ref{secsma} and \ref{secsasbos} is purely a consequence of the different ways of calculating the bosonic Hamiltonian that were used. In the following, we will use the single-mode approximation described in Ref.~\onlinecite{ma90}.

\section{In-plane magnetic field}

Having established the method of choice, let us consider the effect of an additional in-plane magnetic field $B_\pl = B \sin \q$ in the BQHS. \comments{We will ignore the effect on the effective layer thickness.} Let us adopt a coordinate system where $\vc{B}_\pl = B_\pl (0, -1, 0)$.\footnote{This somewhat strange choice of coordinate system is made to facilitate the comparison with Ref.~\onlinecite{ya96}. Note that from this point on, we are again using the spin orientation chosen in Sec. \ref{secsystem}.} The vector potential corresponding to $\vc{B}_\pl$ is $\vc{A}_\pl = B_\pl (0, 0, x)$; thus, a particle tunnelling between the two layers picks up a space-dependent phase. The tunnelling term takes the form~\cite{ya96}
\begin{align}\label{eqhrealspace}
\begin{split}
H_T = & \, - t \int d^2 r \, \vc{h}(\vc{r}) \cdot \vc{S}(\vc{r}), \\
\vc{h}(\vc{r}) = & \, (0,\sin (Qx),\cos (Qx)),
\end{split}
\end{align}
$Q=2\pi B_\parallel d/\phi_0$ is the characteristic momentum introduced by the in-plane field, and $\f_0 = h / e$ is the magnetic flux quantum. In momentum space, $H_T$ is simply
\begin{align} \label{eqHT}
H_T = - \frac{t}{2} \bigl[ S^z_\vc{Q} + i S^y_\vc{Q} + S^z_\vc{-Q} - i S^y_\vc{-Q} \bigr],
\end{align}
where we have written $ \mathbf{Q}=Q\hat x$. From Eq.~\eqref{eqhrealspace}, it is obvious that the tunnelling term favours spin configurations in which the spins align with $\vc{h}$, and thus vary their orientation locally. Allowing for a space-dependent orientation $\f(\vc{r})$ and then calculating the ground state energy yields the Pokrovsky-Talapov model, as shown below.

\subsection{Pokrovsky-Talapov model} \label{secpt}

Let us construct a ground state with locally varying spin orientation,
\begin{align} \label{eqrotspings}
\begin{split}
\ket{\c'} & = \exp \left(i \int d^2r\, S^x(\vc{r}) \f(\vc{r}) \right) \ket{\c} \\
& = \exp \left(i \sum_\vc{q} S^x_\vc{q} \f_\vc{-q}) \right) \ket{\c}.
\end{split}
\end{align}
We will use the shorthand notation $\G = \sum_\vc{q} S^x_\vc{q} \f_\vc{-q}$. The exponential rotates the spin at position $\mathbf{r}$ by an angle $\phi(\mathbf{r})$. It is easy to check that the rotated spin state $\ket{\c'}$ is properly normalised.

We wish to calculate $\bra{\c'} H \ket{\c'}$. Assuming that the spin orientation rotates slowly as a function of position, we can take $\vc{q} \f_\vc{-q}$ to be small. Expanding the exponentials, we obtain a power series in $\f(\vc{r})$. Taking this series up to $\ord(\vc{q}^2 \f^2)$, we obtain for the Coulomb term
\begin{align} \label{eqstiffnessterm}
\bra{\c'} H_C \ket{\c'} = \frac{\r_s}{2} \int d^2 r |\nabla \f(\vc{r})|^2,
\end{align}
where
\begin{align}
\r_s = \frac{1}{32 \p^2} \int k \, dk \, v_E(k) \, e^{-(l k)^2 / 2} (lk)^2,
\end{align}
as found in Ref.~\onlinecite{mo95}. Here, $v_E$ is the interlayer interaction, given by $v_0 - v_c$. For the details of the expansion, see Appendix \ref{appcoulombrot}. Evaluating $\bra{\c'} H_T \ket{\c'}$, with $H_T$ as given in Eq.~\eqref{eqHT}, we find after a similar calculation that
\begin{align} \label{eqcosineterm}
\begin{split}
\bra{\c'} H_T \ket{\c'} = - \frac{t}{2 \p l^2} \int d^2 r\, \cos [\f(\vc{r}) - \vc{Q} \cdot \vc{r}]
\end{split}
\end{align}
(see Appendix \ref{apptunrot} for the details). Combining Eqs.~\eqref{eqcosineterm} and \eqref{eqstiffnessterm}, we obtain the total ground state energy in the presence of tunnelling and an in-plane field:
\begin{align} \label{eqpt}
\begin{split}
E[\f] = & \, \int d^2 r \Bigl\{ \frac{\r_s}{2} |\nabla \f(\vc{r})|^2 - \frac{t}{2 \p l^2} \cos[ \f(\vc{r}) - \vc{Q} \cdot \vc{r} ] \Bigr\}.
\end{split}
\end{align}
This is precisely the Pokrovsky-Talapov model. The Coulomb interaction gives rise to a spin stiffness term: in order to minimise the Coulomb energy, all spins should be aligned. The combination of a nonzero tunnelling amplitude and an in-plane field results in a locally varying preferred spin orientation. In the case of vanishing tunnelling, $t=0$, all the spins will be parallel to each other and aligned in some arbitrary direction in the $yz$-plane. If, on the other hand, $t\neq 0$ but there is no in-plane magnetic field so that $Q=0$, the degeneracy with respect to spin rotations is lifted and the spins will point in the $z$-spin direction.

\subsection{Commensurate-incommensurate transition}

If the spin stiffness is small, it costs little energy to have neighbouring spins with different orientations. In that case, the system will minimise the tunnelling energy by setting $\f(\vc{r}) = \vc{Q} \cdot \vc{r}$, i.e. the spins follow the rotation imposed by the tunnelling term. This is called the \emph{commensurate} phase. If, on the other hand, the spin stiffness is large, the gradient term represents a high energy cost associated with a non-uniform spin orientation and the system will give up the tunnelling energy in favour of a better Coulomb correlation. In that case, the spins do not follow the tunnelling term; this is called the \emph{incommensurate} phase. The onset of this phase is characterised by the appearance of solitons: sudden rotations of the spin orientation by $2 \p$. Fig.~1 shows a sketch of the behaviour of $\f(\vc{r})$ in the incommensurate phase, with two solitons visible.

\begin{figure}
\includegraphics[width=.45\textwidth]{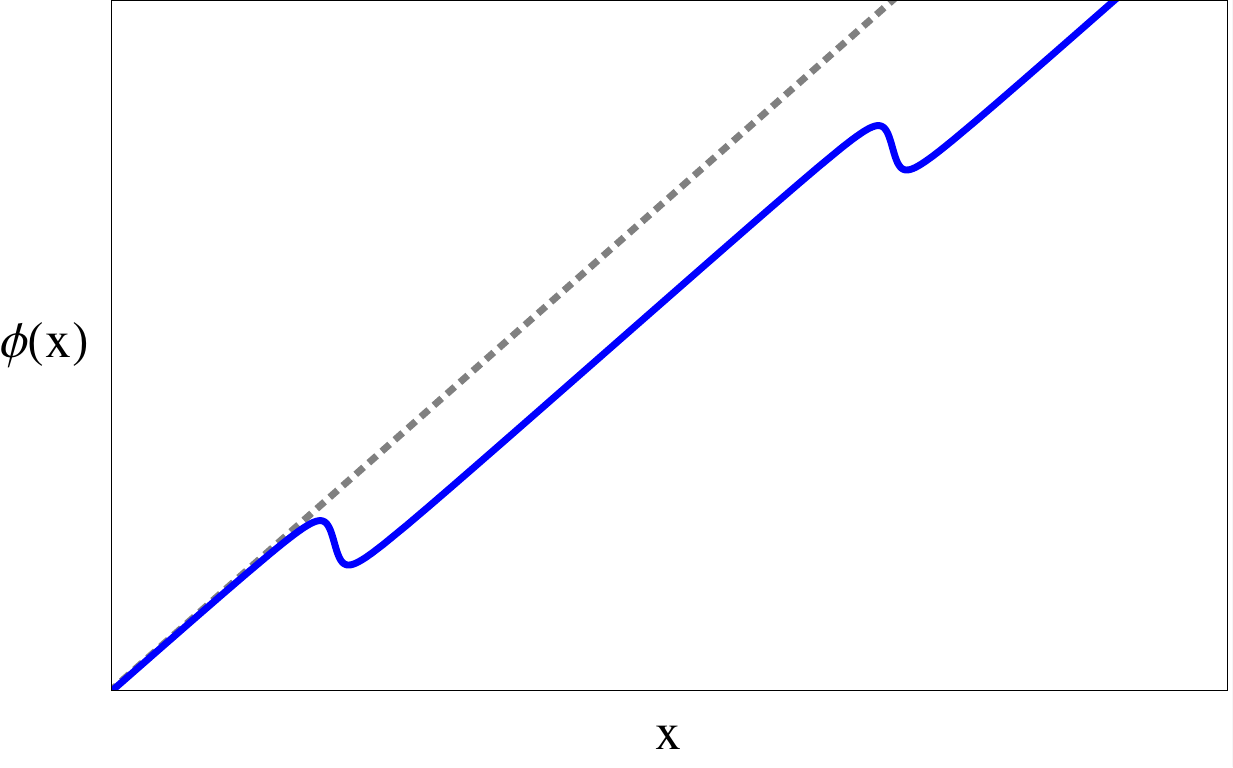}
\caption{(Colour online) Sketch of $\f(x)$ in the incommensurate phase. The gray (dashed) trace indicates $Qx$ while the blue (solid) indicates $\f(x)$. Two solitons (sudden rotations of the phase by $2\pi$) are visible in the figure. In the commensurate phase, $\f$ would exactly follow $Qx$.}
\end{figure}

\subsubsection{Zero temperature}\label{seczerotemp}

The commensurate-incommensurate transition is governed by two quantities: the ratio $t / \r_s$ and the modulus of the characteristic momentum $|l\vc{Q}| = (d/l) \tan \q$. In order to compare the theory to experiments, we need an equation for the critical in-plane field strength. Neglecting finite-temperature effects, we obtain such an equation from the energy functional given in Eq.~\eqref{eqpt}. We consider the energy of a single soliton in the system, i.e. a single rotation by $2 \p$. The phase transition occurs when a finite soliton density is energetically favourable compared to the commensurate (zero-soliton) state. The critical value of $|l\vc{Q}|$ is then given by\cite{alip, ha01}
\begin{align}
\comments{e^{|l\vc{Q}_c|^2 / 8}} |l\vc{Q}_c| = \left( \frac{2}{\p} \right)^{3/2} \sqrt{\frac{t}{\r_s}}.
\end{align}
Recalling that $B_\pl = B_\perp \tan \q$ and that $\tan \q = |l\vc{Q}| \, l / d$, we find
\begin{align}\label{critang}
B_\pl^{\rm c} = B_\perp \frac{l}{d} \left(\frac{2}{\p}\right)^{3/2} \sqrt{\frac{t}{\r_s}}.
\end{align}
Since $\r_s$ depends on $d / l$ and $E_C$, which in turn depend only on $B_\perp$ and constants, this equation predicts the critical in-plane field for a given perpendicular field. In order to compare the theory to the experimental findings presented in Ref.~\onlinecite{fu08}, we need to express the total electron density $n_T$ as a function of the critical in-plane field. The total electron density is related to $B_\perp$ by
\begin{align}
n_T = \n_T n_\f = \frac{1}{2 \p l^2} = \frac{e B_\perp}{2 \p \hbar},
\end{align}
so we can simply invert Eq.~\eqref{critang} to obtain $n_T$ as a function of $B_\pl^c$ for the comparison.

\subsubsection{Finite temperature}\label{secfintemp}

To include the effect of a finite temperature, one has to consider the partition function\cite{alip}
\begin{align}\label{partition}
\mathcal{Z} = \int \mathcal{D} \f \, e^{-E[\f] / k_B T},
\end{align}
where $k_B$ is Boltzmann's constant and $T$ the temperature. \comments{ The quantity of interest is the free energy, which is given by
\begin{align}
F = - k_B T \ln \mathcal{Z}.
\end{align}
}
The functional integral in Eq.~\eqref{partition} can be performed step by step, using the renormalisation group (RG) technique. In every step, an infinitesimal part of the integration is carried out, yielding an effective energy functional at an infinitesimally lower momentum scale, or cut-off. By computing the change in the parameters in the energy functional under an infinitesimal change in the cut-off, one obtains the \emph{flow equations} for $\r_s$ and $t$.\cite{alip} Integrating these equations over the cut-off running from its initial value to zero, one obtains an effective energy functional $E'[\f]$ for which we have
\begin{align}
\mathcal{Z} = e^{-E'[\f] / k_B T}.
\end{align}
$E'$ contains the effective values of $\r_s$ and $t$ at temperature $T$, allowing us to compute the renormalised value of the critical $|l\vc{Q}|$
\begin{align}
|l \vc{Q}_c|(\e) = e^{-\e} \left( \frac{2}{\p} \right)^{3/2} \frac{\sqrt{\r_s(\e) t(\e)}}{\r_s(0)},
\end{align}
where $\e = \ln(\L_0 / \L)$, with $\L_0$ being the initial value of the cut-off $\L$. The RG procedure for the Pokrovsky-Talapov model is discussed in detail in Ref.~\onlinecite{alip}.

\subsection{Comparison with experiment}

In recent experimental work, a commensurate-incommensurate phase transition has been accurately measured.\cite{fu08} Working at a temperature $T = 130$ mK, measuring on a sample with $d = 23$ nm and $t = \D_{\rm SAS} / 2 = 5.5$ K, the red dots in Fig.~2 were obtained. The parameter values reported in Ref.~\onlinecite{fu08} give a $|l\vc{Q}_c|$ of about 2.1-2.2. This value invalidates the assumption that $|l\vc{Q}| \ll 1$, which was made to derive the Pokrovsky-Talapov model. Nonetheless, it turns out that Eq.~\eqref{critang} predicts the experimentally observed critical in-plane fields in the regime of low electron densities, as can be seen in Fig.~2. As the electron density increases, the accuracy of the prediction decreases. This may be related to the fact that increasing the electron density is equivalent to decreasing the magnetic length $l$, and hence, to increasing the effective interlayer separation $d / l$, whereas the dependence of $\r_s$ on $d / l$ was derived under the assumption that $d / l$ is small.

Including finite-temperature effects in the manner described in Sec.~\ref{secfintemp} reduces the predicted critical in-plane field (see Ref.~\onlinecite{alip}). At the temperature reported in Ref.~\onlinecite{fu08}, we find a very small change in the critical in-plane field strength (see Fig.~2, green trace). On the other hand, the smallness of the shift validates the assumptions made during the experimental data analysis.\cite{kup}

Testing the theory on the experimental data provided by Murphy \emph{et al.}\cite{mu94}, who measured on a sample with $n_T = 1.26 \times 10^{15}$ m$^{-2}$, $t = 0.4$ K and $d = 21$ nm, we find a critical angle of inclination of $18^\circ$, about a factor of 2 off from the reported value of $8 \pm 2 ^\circ$, as already found by Kun Yang \emph{et al.}.\cite{ya96} Taking finite-temperature ($T = 0.4$ K) effects into account, this prediction improves to $15.7^\circ$: again a small change in the right direction.

\begin{figure} \label{figphasediag}
\includegraphics[width=.5\textwidth]{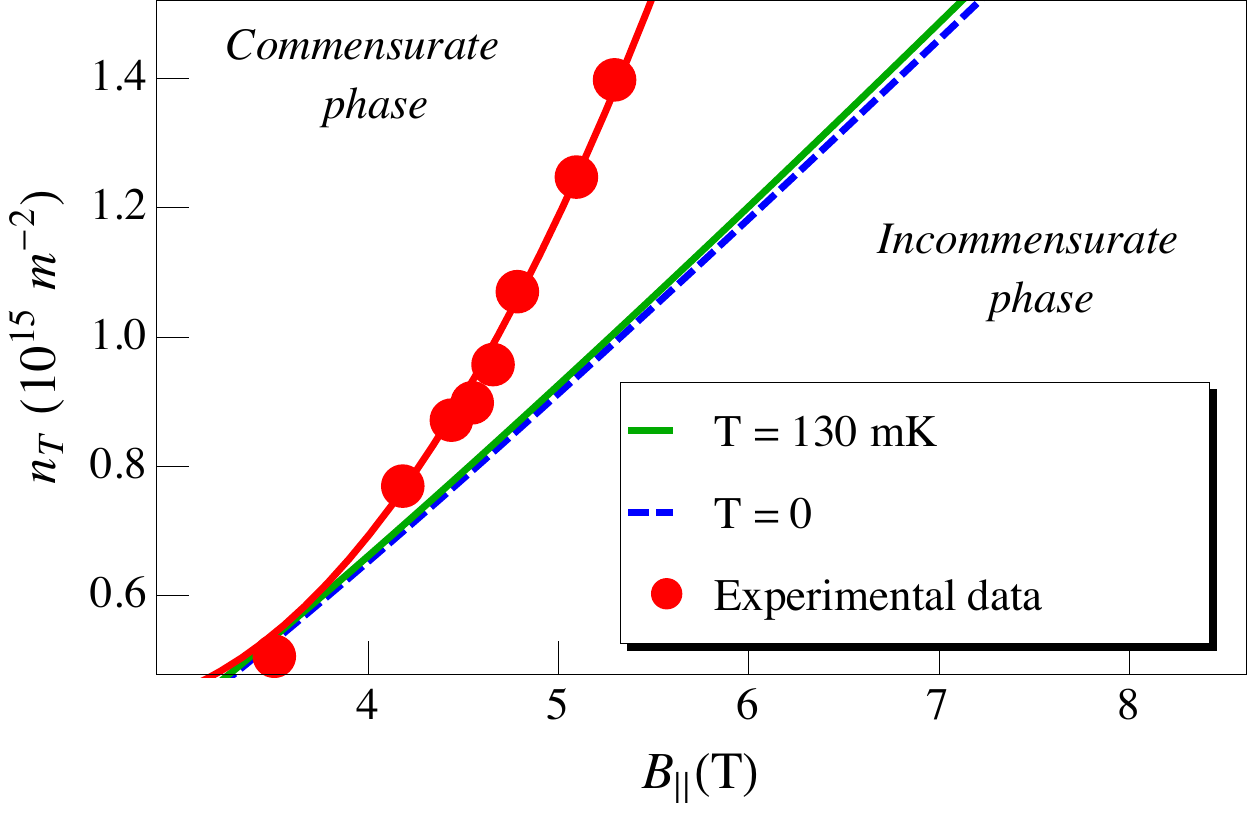}
\caption{(Colour online) In-plane field vs.\ total electron density phase diagram indicating the commensurate-incommensurate phase transition. The experimental data are given by the red dots (the trace is a guide for the eye). The theoretical predictions are given by the blue, dashed line (mean field) and the green, solid line (finite temperature). The experimental data are taken from Ref.~\onlinecite{fu08}.}
\end{figure}

\section{Discussion \& conclusions}
In this paper, our aim was twofold: firstly, to clarify the situation of the many magnon models for a BQHS at $\n_t = 1$; and secondly, to study the effects of an in-plane field in such a system. We have considered two models for a BQHS at zero temperature and $\n_t = 1$, both of which turn out to be variations of the magnon model obtained by expanding the Hamiltonian in $n$-magnon states. We have also seen what the origin of the difference between the two models is: it is the moment in the calculation at which the magnons are approximated to be bosonic.

We have derived the Pokrovsky-Talapov model for the ground state energy of a bilayer QHS at $\n_t = 1$ with an in-plane field from the underlying microscopic physics, which had been suggested earlier by Kun Yang \emph{et al}.\cite{ya94} We have seen the phases predicted by this model in the weak in-plane field limit: a commensurate phase, in which the pseudospin follows the underlying structure provided by the combination of the tunnelling term and in-plane field, and an incommensurate phase, where the pseudospin rotates incommensurately with the underlying structure. We observed that this model, although derived for weak in-plane fields, predicts experimentally measured values in the regime of low electron densities, and we have offered an explanation for the loss of agreement at higher electron densities. We have showed that the inclusion of finite-temperature effects produces only small effects at the temperatures reported in Refs.~\onlinecite{fu08, mu94}. Even though the effects considered here are small, a finite-temperature-adjusted theory is a potentially valuable asset in future research.

\section*{Acknowledgements}
We acknowledge financial support from the Netherlands Organization for Scientific Research (NWO). We thank L.K. Lim for stimulating discussions and R.L. Doretto for sharing his expertise on the subject. Furthermore, we express our gratitude to A. Fukuda and N. Kumada for sending us the details of their experimental setup.

\appendix

\section{The LLL projection algebra}
\label{lllprojectionalg}

The function $G_{m, n}(\vc{k})$ is given by\cite{do04}
\begin{widetext}
\begin{align}
G_{m, n}(\vc{k}) = & \, \q(m - n) \sqrt{\frac{m!}{n!}} \left( \frac{-i l \tl{k}^*}{\sqrt{2}} \right)^{n - m} L^{n - m}_m \left( \frac{|l\vc{k}|^2}{2} \right) + \q(n - m) \sqrt{\frac{n!}{m!}} \left( \frac{-i l \tl{k}}{\sqrt{2}} \right)^{m - n} L^{m - n}_m \left( \frac{|l\vc{k}|^2}{2} \right),
\end{align}
\end{widetext}
where $\tl{k} = k_x + i k_y$. Making use of the Landau level basis, we can calculate the sum of the product of two $G_{m, n}$-functions (i.e., the matrix product $G(\vc{k}) G(\vc{q})$). We obtain
\begin{align}\label{ggprod}
\sum_l G_{m, l}(\vc{k}) G_{l, n}(\vc{q}) = e^{-l^2 \vc{k} \cdot \vc{q} / 2} e^{- i \vc{k} \wedge \vc{q} / 2} G_{m, n}(\vc{k} +
\vc{q})
\end{align}
where $\vc{k} \wedge \vc{q} = l^2 \hat{z} \cdot (\vc{k} \times \vc{q})$. Eq.~\eqref{ggprod} gives rise to the LLL projection algebra, which can be summarised as follows:
\begin{align} \label{eqlllprjalg1}
\begin{split}
[\r_\vc{k}, \r_\vc{q}] = & \, e^{l^2 \vc{k} \cdot \vc{q} / 2} 2 i \sin(\vc{k} \wedge \vc{q} / 2) \r_{\vc{k} + \vc{q}} \\
[S^\m_\vc{k}, \r_\vc{q}] = & \, e^{l^2 \vc{k} \cdot \vc{q} / 2} 2 i \sin(\vc{k} \wedge \vc{q} / 2) S^\m_{\vc{k} + \vc{q}} \\
[S^\m_\vc{k}, S^\n_\vc{q}] = & \, e^{l^2 \vc{k} \cdot \vc{q} / 2} \Bigl( \frac{i \d_{\m, \n}}{2} \sin(\vc{k} \wedge \vc{q} / 2)
\r_{\vc{k} + \vc{q}} \\
& \qquad \qquad + i \e^{\m \n \s} \cos(\vc{k} \wedge \vc{q} / 2) S^\s_{\vc{k} + \vc{q}} \Bigr).
\end{split}
\end{align}
We can also derive the following commutators, for later use:
\begin{align} \label{eqlllprjalg2}
\begin{split}
[S^+_\vc{k}, S^-_\vc{q}] = & \, e^{l^2 \vc{k} \cdot \vc{q} / 2} \Bigl( i \sin(\vc{k} \wedge \vc{q} / 2) \r_{\vc{k} + \vc{q}} \\
& \qquad \qquad + 2 \cos(\vc{k} \wedge \vc{q} / 2) S^z_{\vc{k} + \vc{q}} \Bigr), \\
[S^\pm_\vc{k}, \r_\vc{q}] = & \, e^{l^2 \vc{k} \cdot \vc{q} / 2} 2 i \sin(\vc{k} \wedge \vc{q} / 2) S^\pm_{\vc{k} + \vc{q}} \\
[S^\pm_\vc{k}, S^z_\vc{q}] = & \, \pm e^{l^2 \vc{k} \cdot \vc{q} / 2} \cos(\vc{k} \wedge \vc{q} / 2) S^\pm_{\vc{k} + \vc{q}}.
\end{split}
\end{align}

\section{Expectation values in the commensurate (rotating-spin) ground state} 
\label{appvevrotspin}

We are interested in expectation values of the form $\bra{\c'} A_\vc{k} \ket{\c'}$, where $\ket{\c'}$ is the rotating-spin ground state defined in Sec.~\ref{secpt}. Hence, we need to evaluate terms of the form $e^{-i \G} A_\vc{k} e^{i \G}$. To this end, we use a simple generalization of the Baker-Hausdorff lemma, and find
\begin{align}
e^{-i \G} A_\vc{k} e^{i \G} = \sum_n C_n(A_\vc{k})
\end{align}
where
\begin{align}
\begin{split}
C_n(A_\vc{k}) = & \frac{1}{n!} \sum_{\vc{q}_1, \dots, \vc{q}_n} (i \f_{-\vc{q}_1}) \dots (i \f_{-\vc{q}_n}) \\
& \times [S^x_{\vc{q}_1}, \dots, [S^x_{\vc{q}_n}, A_\vc{k}] \dots ].
\end{split}
\end{align}

\subsection{Coulomb term} \label{appcoulombrot}

Using Eq.~\eqref{eqlllprjalg1}, it is easy to verify that for odd $n$, $C_n(\r_\vc{k}) \propto S^x_{\vc{k} + \vc{q}_1 + \dots + \vc{q}_n}$, whereas for even $n$, $C_n(\r_\vc{k}) \propto \r_{\vc{k} + \vc{q}_1 + \dots + \vc{q}_n}$. For $C_n(S^x_\vc{k})$, we obtain the same solutions, but even and odd $n$ are inverted with respect to $C_n(\r_\vc{k})$. Hence, projecting the Coulomb term $H_C$ into the spin-rotating ground state $\ket{\c'}$, we obtain
\begin{align*}
\bra{\c'} H_C \ket{\c'} = \frac{1}{2} & \, \sum_{\vc{k}, n, n'} v_0(\vc{k}) \bra{\c} C_n(\r_\vc{-k}) C_{n'}(\r_\vc{k}) \ket{\c} \\
+ & \, 2 \sum_{\vc{k}, n, n'} v_c(\vc{k}) \bra{\c'} C_n(S^x_\vc{-k}) C_{n'}(S^x_\vc{k}) \ket{\c'}.
\end{align*}
In Ref.~\onlinecite{mo95}, this series is taken up to $n + n' = 2$. In that case, with the assumption of a slowly varying spin orientation, one obtains the result presented in Eq.~\eqref{eqstiffnessterm}.

\begin{widetext}

\subsection{Tunnelling term} \label{apptunrot}

The tunnelling term $H_T$ can be analysed in the same way. We need
\begin{align} \label{eqvevhtrot}
\bra{\c'} H_T \ket{\c'} = - \frac{t}{2} \sum_n \bra{\c_0} \Bigl( & C_n(S^z_\vc{Q}) + C_n(S^z_{-\vc{Q}}) + i C_n(S^y_\vc{Q}) - i C_n(S^y_{-\vc{Q}}) \Bigr) \ket{\c_0}.
\end{align}
From Eq.~\eqref{eqlllprjalg1}, we can deduce that the surviving terms are of the form $C_{2n}(S^z_\vc{k})$ and $i \, C_{2n + 1}(S^y_\vc{k})$. Repeated application of Eq.~\eqref{eqlllprjalg1} yields
\begin{align} \label{eqc2n}
\begin{split}
C_{2n}(S^z_\vc{Q}) = & \, \frac{(-1)^n}{(2n)!} \sum_{\vc{q}_1, \dots \vc{q}_{2n}} \f_{-\vc{q}_1} \dots \f_{-\vc{q}_{2n}} e^{l^2 \vc{q}_{2n} \cdot \vc{Q} / 2} \dots e^{l^2 \vc{q}_1 \cdot (\vc{q}_2 + \dots + \vc{Q}) / 2} \\
& \, \times \cos(\vc{q}_{2n} \wedge \vc{Q} / 2) \dots \cos(\vc{q}_1 \wedge (\vc{q}_2 + \dots + \vc{Q}) / 2) S^z_{\vc{Q} + \vc{q}_1 + \dots + \vc{q}_{2n}}.
\end{split}
\end{align}
and similarly for $C_{2n + 1}(S^y_\vc{Q})$. Since we are assuming a slow modulation, we are working at small $l \vc{Q}$, and we may take $l^2 |\vc{Q}||\vc{q}| \approx 0$. Evaluating $S^z_{\vc{Q} + \vc{q}_1 + \dots + \vc{q}_{2n}}$ in the uniform ground state gives a Kronecker delta; inserting this into the sum from Eq.~\eqref{eqc2n} and Fourier transforming back to real space, we find
\begin{align}
\sum_n \bra{\c_0} C_n(S^z_\vc{Q}) \ket{\c_0} \notag = N \int d^2 r \, e^{i \vc{Q} \cdot \vc{r}} \sum_n \frac{(-1)^n}{(2n)!} \f^{2n}(\vc{r}).
\end{align}
The sum is an expansion in powers of $\f$ of the function $\cos(\f(\vc{r}))$. Similar calculations for the other three terms from Eq.~\eqref{eqvevhtrot} yield
\begin{align}
\bra{\c_0'} H_T \ket{\c_0'} = & \, - N \frac{t}{2} \int d^2 r \Bigl\{ e^{i \vc{Q} \cdot \vc{r}} \Bigl[ \cos(\f(\vc{r})) - i \sin(\f(\vc{r})) \Bigr] + e^{-i \vc{Q} \cdot \vc{r}} \Bigl[ \cos(\f(\vc{r})) + i \sin(\f(\vc{r})) \Bigr] \Bigr\} \notag \\
= & \, - \frac{t}{2 \p l^2} \int d^2 r \, \cos(\f(\vc{r}) - \vc{Q} \cdot \vc{r}).
\end{align}
This equation appeared first in Ref.~\onlinecite{ya94}, but since it was not derived there, we include it here.
\end{widetext}


\begin{thebibliography}{00}
\bibitem{ke04} M. Kellogg, J.P. Eisenstein, L.N. Pfeiffer, and K.W. West, Phys. Rev. Lett. {\bf 93}, 036801 (2004).

\bibitem{ke02} M. Kellogg, I.B. Spielman, J.P. Eisenstein, L.N. Pfeiffer, and K.W. West, Phys. Rev. Lett. {\bf 88}, 126804 (2002).

\bibitem{tu03} E. Tutuc, M. Shayegan, and D.A. Huse, Phys. Rev. Lett. {\bf 93}, 036802 (2004).

\bibitem{sp01} I.B. Spielman, J.P. Eisenstein, L.N. Pfeiffer, and K.W. West, Phys. Rev. Lett. {\bf 87}, 036803 (2001).

\bibitem{ch08} A.R. Champagne, A.D.K. Finck, J.P. Eisenstein, L.N. Pfeiffer, and K.W. West, Phys. Rev. B {\bf 78}, 205310 (2008).

\bibitem{mu94} S.Q. Murphy, J.P. Eisenstein, G.S. Boebinger, L.N. Pfeiffer, and K.W. West, Phys. Rev. Lett. {\bf 72}, 728 (1994).

\bibitem{ya94} K. Yang, K. Moon, L. Zheng, A.H. MacDonald, S.M. Girvin, D. Yoshioka, and Shou-Cheng Zhang, Phys. Rev. Lett. \textbf{72}, 732 (1994).

\bibitem{fu08} A. Fukuda, D. Terasawa, M. Morino, K. Iwata, S. Kozumi, N. Kumada, Y. Hirayama, Z.F. Ezawa, and A. Sawada, Phys. Rev. Lett. {\bf 100}, 016801 (2008).

\bibitem{fe89} H. A. Fertig, Phys. Rev. B {\bf 40}, 1087 (1989).

\bibitem{ma90} A. H. MacDonald, P.M. Platzman, and G.S. Boebinger, Phys. Rev. Lett. {\bf 65}, 775 (1990).

\bibitem{do06} R. L. Doretto, A. O. Caldeira, and C.M. Smith, Phys. Rev. Lett. {\bf 97}, 186401 (2006).

\bibitem{ya01} K. Yang, Phys. Rev. Lett. {\bf 87}, 056802 (2001)

\bibitem{ka09} B. Karmakar, V. Pellegrini, A. Pinczuk, L.N. Pfeiffer, and K.W. West, Phys. Rev. Lett. {\bf 102}, 036802 (2009).

\bibitem{ya96} K. Yang, K. Moon, Lotfi Belkhir, H. Mori, S.M. Girvin, and A.H. MacDonald, L. Zheng, and D. Yoshioka, Phys. Rev. B {\bf 54}, 11644 (1996).

\bibitem{ha01} C.B. Hanna, A.H. MacDonald, and S.M. Girvin, Phys. Rev. B {\bf 63}, 125305 (2001).

\bibitem{alip} A. Lazarides, O. Tieleman, and C. Morais Smith, arXiv 0909:0526.

\bibitem{gi86} S.M. Girvin, A.H. MacDonald, and P.M. Platzman, Phys. Rev. B {\bf 33}, 2481 (1986).

\bibitem{la80} L. Landau and E. Lifshitz, Statistical Physics Part 2, Course of Theoretical Physics Vol. 9 (Pergamon Press, 1980).

\bibitem{eispriv} J.P. Eisenstein, private communcation.

\bibitem{do04} R. L. Doretto, A. O. Caldeira, and S.M. Girvin, Phys. Rev. B {\bf 71}, 045339 (2005).

\bibitem{mo95} K. Moon, H. Mori, K. Yang, S.M. Girvin, A.H. MacDonald, L. Zheng, D. Yoshioka, and Shou-Cheng Zhang, Phys. Rev. B {\bf 51}, 5138 (1995).

\bibitem{kup} N. Kumada, private communication.

\end{thebibliography}
\end{document}